\documentclass{aastex631}
\usepackage{amsmath}
\usepackage{soul}

\newcommand{\Ub}{{\bf U}}

\shorttitle{Neural Networks for Nuclear Reactions in MAESTROeX}
\shortauthors{Fan et al.}

\begin{document}

\title{Neural Networks for Nuclear Reactions in MAESTROeX}

\correspondingauthor{Andrew Nonaka}
\email{AJNonaka@lbl.gov}

\author[0000-0002-3246-4315]{Duoming Fan}
\affiliation{Lawrence Berkeley National Laboratory,
Center for Computational Sciences and Engineering,
One Cyclotron Road, MS 50A-3111, Berkeley, CA 94720, USA}

\author[0000-0003-2300-5165]{Donald E. Willcox}
\affiliation{Lawrence Berkeley National Laboratory,
Center for Computational Sciences and Engineering,
One Cyclotron Road, MS 50A-3111, Berkeley, CA 94720, USA}

\author[0000-0002-7815-1496]{Christopher DeGrendele}
\affiliation{University of California Santa Cruz,
Applied Mathematics Department,
1156 High Street, Santa Cruz, CA 95060, USA}

\author[0000-0001-8401-030X]{Michael Zingale}
\affiliation{Stony Brook University,
Department of Physics and Astronomy,
Stony Brook, NY 11794-3800}


\author[0000-0003-1791-0265]{Andrew Nonaka}
\affiliation{Lawrence Berkeley National Laboratory,
Center for Computational Sciences and Engineering,
One Cyclotron Road, MS 50A-3111, Berkeley, CA 94720, USA}



\begin{abstract}
We demonstrate the use of neural networks to accelerate the reaction steps in the MAESTROeX stellar hydrodynamics code.
A traditional MAESTROeX simulation uses a stiff ODE integrator for the reactions;
here we employ a ResNet architecture and describe details relating to the architecture, training, and validation of our networks.
Our customized approach includes options for the form of the loss functions, a demonstration that the use of parallel neural networks leads to increased accuracy, and a description of a perturbational approach in the training step that robustifies the model.
We test our approach on millimeter-scale flames using a single-step, 3-isotope network describing the first stages of carbon fusion occurring in Type Ia supernovae.
We train the neural networks using simulation data from a standard MAESTROeX simulation, and show that the resulting model can be effectively applied to different flame configurations.
This work lays the groundwork for more complex networks, and iterative time-integration strategies that can leverage the efficiency of the neural networks.\\ \\
{\it Unified Astronomy Thesaurus concepts:}
Neural networks (1933);
Reaction models (2231);
Nucleosynthesis (1131);
Computational methods (1965);
Hydrodynamical simulations (767)
\end{abstract}



\section{Introduction} \label{sec:Introduction}
In stellar astrophysics simulations, nuclear reactions are often the most computationally demanding aspect.  Even in moderately complex networks, the time scales of the stiffest reactions can be on the order of pico- or even femtoseconds.
In explicit reaction integration schemes, this leads to an overly restrictive (compared to advective, acoustic, or diffusive scale) time step, and for implicit schemes, can require hundreds or thousands of evaluations of the rates per time step.
Thus, reactions can be orders of magnitude more expensive than advection and/or implicit global solvers such as self-gravity, momentum, or mass diffusion.
Even with a relatively simple three-species, one-step reaction network, implicit treatment of burning can require nearly half of the total computational time.

One burgeoning approach to computational fluid dynamics (CFD) is the use of machine learning approaches to replace various computational kernels such as advection~\citep{papapicco2021neural}, diffusion~\citep{sirignano2018dgm}, and the Poisson equation~\citep{tang2017study} (for self-gravity, electrostatics, or projection-based decomposition techniques for incompressible flow).
In particular, the use of Deep Neural Networks (DNNs) for surrogate modeling in the framework of partial differential equations (PDEs) and ordinary differential equations (ODEs) has been especially popular in recent years. However, while literature on using DNN models in areas of CFD such as turbulence modeling is increasing rapidly~\citep{echekki2015turbcomb, duraisamy2019turbreview, grimberg2020cfdturb, lye2020deeplearncfd}, there is notably less investigations in using them for chemical kinetics; see \cite{meuwly2021machine} for a review of current machine learning techniques for chemical reactions in a wide range of applications.
Indeed, more interest in using neural networks for computational chemistry kernels in reacting flow simulations have only very recently been shown, with several groups having used this approach on moderate-sized systems ($\sim$10 species, $\sim$20 reactions) in a terrestrial combustion context~\citep{brown2021novel,owoyele2020chemnode,sharma2020deep,ji2020stiff}.  Astrophysical nuclear reaction networks share a lot in common with terrestrial chemical networks, although the astrophysical rates tend to have much stronger temperature dependence, and hence can be more stiff.

In the context of using a DNN to accelerate or replace computationally-expensive PDE or ODE solve, the DNN model is trained to approximate the mapping from the input states of the differential equation solver to its solution states. The learning problem then aims at tuning the weights of the DNN model to train it to a high degree of accuracy. Examples of DNN for surrogate modeling include Residual Neural Network (ResNet)~\citep{he2016deep}, physics-informed neural network (PINN)~\citep{karniadakis2021physics,raissi2019physics}, and fractional DNN~\citep{antil2020fracdnn}. Among them, PINN has achieved success in a wide range of applications by encoding physics constraints into the loss functions of the neural network such that the governing equations are satisfied. However, recent investigations by \cite{ji2020stiff} and \cite{wang2020stiffpinn} have shown that the performance of using PINN in stiff chemical kinetic problems with governing equations of stiff ODEs is sub-optimal and can often fail due to both numerical stiffness and physical stiffness. Hence, our approach to constructing a surrogate model for the reaction computational kernel is more similar to \cite{brown2021novel} where the goal is to {\em Learn-from-Physics/Chemistry}. In addition, we still want to include physics-based constraints in the loss function whenever possible, but they will not be expressed in the form of the governing equations and instead be specific to the problem (for example, conservation of mass).

In this paper we use the astrophysical hydrodynamics code, MAESTROeX \citep{nonaka2010maestro,fan2019maestroex}, to perform millimeter-scale nuclear flame simulations using a neural network to accelerate the reaction steps.
We demonstrate that we can train neural networks using data from a traditional MAESTROeX simulation that utilizes stiff ODE integration for the reactions, and run new simulations utilizing this neural network at a reduced computational cost.
This particular problem is extremely challenging due to the delicate balance between advection, thermal diffusion, and reactions.
We have implemented a training scheme which is highly accurate and sensitive to this balance, and describe our design decisions below.



\section{Model and Prior Numerical Approach}\label{sec:Model}

MAESTROeX is a finite-volume code that solves the equations of low Mach number reacting flow in astrophysical environments.
Since the low Mach number model does not contain acoustic waves, the time step is limited by an advective CFL constraint, which is $\mathcal{O}(1/{\rm Ma})$ larger than an acoustic CFL constraint in compressible approaches, where Ma is the characteristic Mach number.
The code is suitable for both stratified environments (planar regions or full stars) as well as small scale simulations without stratification (where it reduces to the system described in \cite{Bell:2004}).
Here we focus on the latter case (a millimeter-scale flame); thus we do not account for gravitational stratification and the model is simpler than the full MAESTROeX equation set,
\begin{eqnarray}
\frac{\partial\rho X_k}{\partial t} &=& -\nabla\cdot(\rho X_k\Ub) + \rho\dot\omega_k,\\
\frac{\partial\rho h}{\partial t} &=& -\nabla\cdot(\rho h\Ub) + \nabla\cdot k_{\rm th}\nabla T + \rho H_{\rm nuc},\\
\frac{\partial\Ub}{\partial t} &=& -\Ub\cdot\nabla\Ub - \frac{1}{\rho}\nabla\pi,\\
\nabla\cdot\Ub &=& S.
\end{eqnarray}
Here $\rho$ is the fluid density, $X_k$ is the mass fraction of species $k$ with associated production rate $\dot\omega_k$, $\Ub$ is the fluid velocity, $h$ is the enthalpy per unit mass, $k_{\rm th}$ is the thermal conductivity, $T$ is the temperature, $H_{\rm nuc}$ is the nuclear energy generation rate per unit mass, and $\pi$ is the perturbational (or dynamic) pressure.
The divergence constraint is derived directly from the equation of state by taking the Lagrangian derivative and substituting in the equations of mass and energy evolution.
The constraint represents a linearized approximation of the velocity field required so that the thermodynamic variables evolve such that $p(\rho,h,{\bf X}) = p_0$, where $p_0$ is a constant ambient pressure.
The expansion term, $S$ accounts for local compressibility effects resulting from nuclear burning, compositional changes, and thermal conduction; see \cite{fan2019maestroex}. 
Millimeter scale flame instabilities with this algorithm have previously been studied in \citet{SNld,SNrt,SNrt3d}.

The MAESTROeX algorithm utilizes a projection methodology for the velocity integration, and Strang splitting for the thermodynamic variable integration.  The projection algorithm involves an explicit hydrodynamic step followed by a global geometric multigrid Poisson solve to correct the solution so that it satisfies the divergence constraint.  The Strang splitting algorithm alternates between a reaction half-step, an advection-diffusion full step, and a reaction half-step to achieve second-order accuracy.  Advection is treated explicitly with a second-order accurate Godunov approach based on the corner transport upwind scheme of \citet{colella:1990}, and thermal diffusion is treated implicitly using a global Helmholtz linear solver using geometric multigrid.
Reactions are integrated using the stiff ODE solver VODE \citep{brown1989vode}.

We use a publicly available equation of state \citep{timmes2000accuracy}, which includes contributions from electrons, ions, and radiation.
We select a simple 3-isotope network describing the first stages of carbon fusion occurring in Type Ia supernovae (SNIa). In simmering and deflagration stages of common SNIa models, $^{12}$C nuclei fuse with other $^{12}$C nuclei to form predominantly either $^{4}$He + $^{20}$Ne or $^{1}$H + $^{23}$Na. Because both of these sets of reaction products can be thought to proceed from the decay of a short-lived $^{24}$Mg nucleus in an excited state, our network describes these reactions as a single $^{12}$C + $^{12}$C $\rightarrow$ $^{24}$Mg reaction to reduce the number of equations and isotopes we track. We evolve $^{12}$C and $^{24}$Mg mass fractions using screening as described in \cite{graboske1973screening,weaver1978presupernova,alastuey1978nuclear,itoh1979enhancement}.
This particular network contains time scales on the order of 10$^{-14}$~s so implicit integration with VODE is required when using large timesteps afforded by the MAESTROeX algorithm.
Because common SNIa models also generally contain about 50\% $^{16}$O by mass in the progenitor white dwarf, we include this quantity of $^{16}$O in our network, comprising the third species in our network. However, we do not evolve $^{16}$O in the network as we are only interested in modeling the early stages of $^{12}$C fusion in SNIa models and have not extended the current study to account for later $^{16}$O burning.


\section{Deep Neural Network}\label{sec:Methods}

\subsection{Problem Formulation}\label{sec:problem_statement}

As stated in Section \ref{sec:Model}, the MAESTROeX algorithm integrates the reactions using the stiff ODE solver VODE, which evolves the species and enthalpy from $X_k^{\rm in} \rightarrow X_k^{\rm out}$ and $(\rho h)^{\rm in} \rightarrow (\rho h)^{\rm out}$ by solving the following system of equations over a time interval of $\Delta t$,
\begin{equation}\label{eq:pde_Xk}
\frac{\partial\rho X_k}{\partial t} = \rho\dot\omega_k, 
\end{equation}
\begin{equation}\label{eq:pde_h}
\frac{\partial(\rho h)}{\partial t} = \rho H_{\rm nuc},
\end{equation}
where in the absence of weak interactions, the nuclear energy generation rate can be expressed in terms of the specific binding energies, $q_k$, as
\begin{equation}
H_{\rm nuc} = -\sum_k q_k\dot{\omega}_k.
\end{equation}
We note that in \cite{fan2019maestroex} we integrated temperature rather than enthalpy; it was later shown in \cite{zingale2021practical} that integrating energy is a more robust approach, which we have subsequently adopted.
In the Strang-split time integration algorithm for MAESTROeX, we evolve only Equations (\ref{eq:pde_Xk}) and (\ref{eq:pde_h}) in our reaction step (not density as well). As a result, as implemented within our reactions module, we integrate internal energy, $e$, rather than enthalpy $h$, since the time derivatives of these quantities from reactions alone are identical, if the pressure is constant in time, which it is for smallscale flames in an open domain.

We seek to replace the current reaction solver using deep neural networks (DNNs) such that the updated mass fractions and enthalpy are determined by ``black-boxes" that satisfy the following: 
\begin{eqnarray}
{\bf X}^{\rm out} &=& \text{DNN}_1 ({\bf X}^{\rm in}, \rho^{\rm in}, T^{\rm in}, \Delta t), \label{eq:dnn_Xk} \\
(\rho h)^{\rm out} &=& (\rho h)^{\rm in} + \rho^{\rm out} \text{DNN}_2 ({\bf X}^{\rm in}, \rho^{\rm in}, T^{\rm in}, \Delta t). \label{eq:dnn_rhoh}
\end{eqnarray}

\noindent Note that the inputs to both DNNs are the same, which allows the two models to be combined during runtime.
Also note that the density remains unchanged during this reaction step, hence the enthalpy update can be expressed simply as 
\begin{equation}\label{eq:dnn_h}
h^{\rm out} = h^{\rm in} + \text{DNN}_2 ({\bf X}^{\rm in}, \rho^{\rm in}, T^{\rm in}, \Delta t).
\end{equation}
Combining Equations (\ref{eq:dnn_Xk}) and (\ref{eq:dnn_h}) into a single neural network gives
\begin{equation}\label{eq:dnn_full}
[{\bf X},h]^{\rm out} = [{\bf X},h]^{\rm in} + \text{DNN}_{\rm full} ({\bf X}^{\rm in}, \rho^{\rm in}, T^{\rm in}, \Delta t)
\end{equation}

\subsection{Architecture}\label{sec:architecture}

To accurately represent the functionality of the ODE solver VODE, the inputs of the DNNs are density, temperature, and mass fractions.
Note that in this work, we use a constant time step in each of our simulations, and hence $\Delta t$ is not used as an input.
For more general problems, $\Delta t$ will also need to be considered.
The input dimension is then given by
\begin{equation}
n_0 = \underset{\rm (density)}{1} + \underset{\rm (temperature)}{1} + \underset{\text{(\# of species)}}{M} = M + 2    
\end{equation}
where $M=2$ for our reaction network as discussed in Section \ref{sec:Model}. Due to the sensitivity of
the nuclear energy generation (expressed in erg/g),
$e_{\rm enuc}$, to the accuracy of the species mass fractions (refer to Appendix \ref{app:enuc_analysis} for a detailed discussion), the output will include both the species mass fractions and nuclear energy generation $e_{\rm nuc}$ for a total output dimension of $M+1$.

We consider two different types of neural networks for this problem: one single neural network DNN$_{\rm full}$ satisfying Equation (\ref{eq:dnn_full}), and two parallel neural networks DNN$_1$ and DNN$_2$ satisfying Equations (\ref{eq:dnn_Xk}) and (\ref{eq:dnn_h}) respectively, which will be combined during run time to approximate the entire solution. It is shown in Section \ref{sec:Results} that the single neural network does not seem to perform as well as the parallel network.  Additional advantages of training the networks in parallel include decreasing the total training time and avoiding potential memory limitations.

\begin{figure}
\gridline{\fig{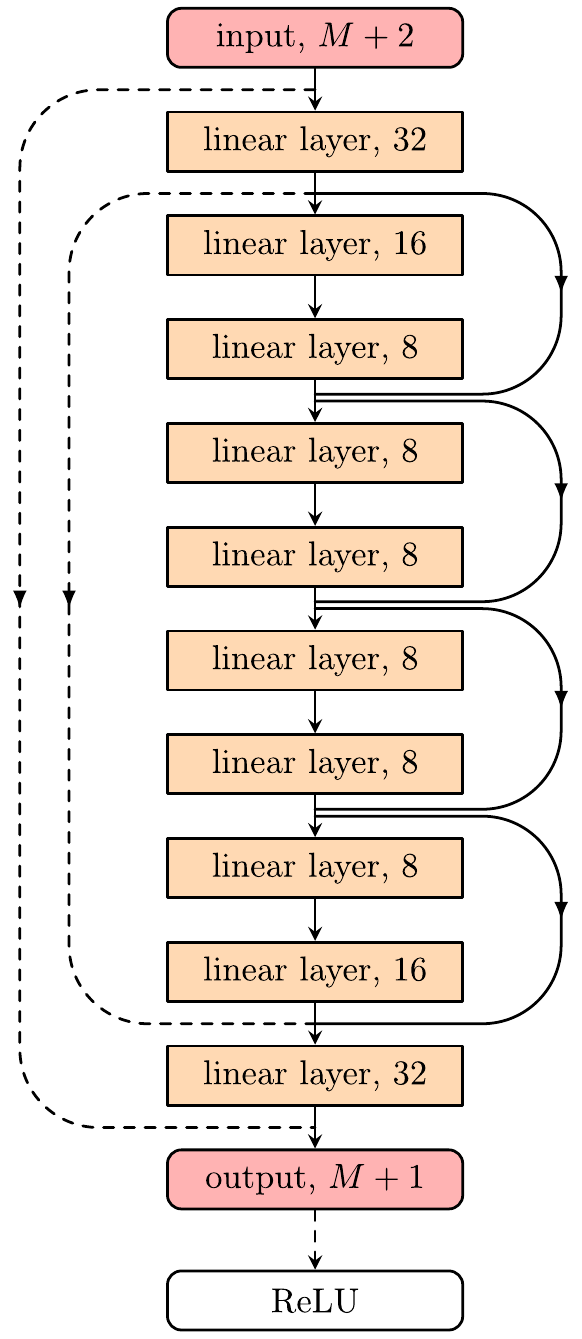}{0.32\textwidth}{(a) DNN$_{\rm full}$}
          \fig{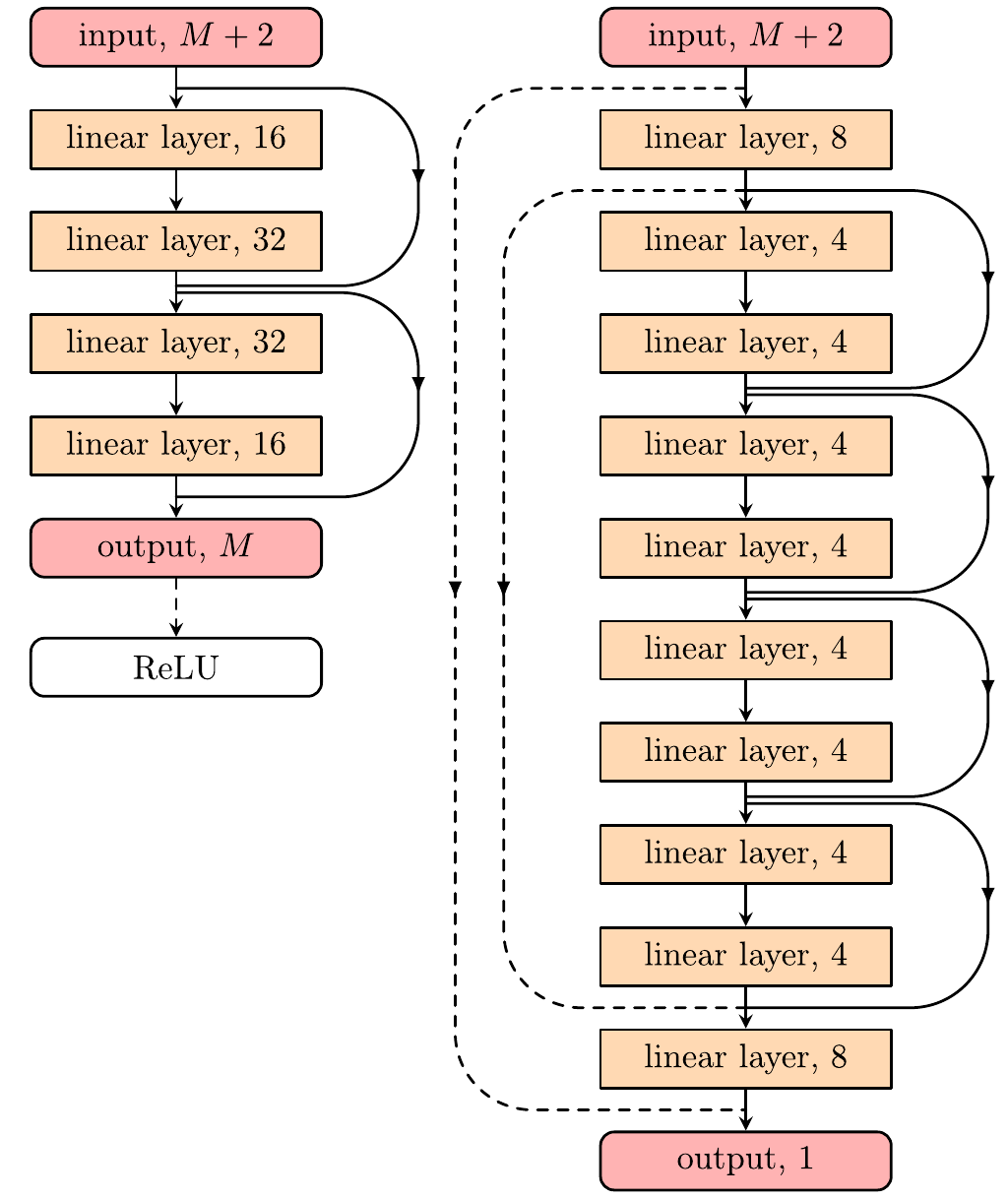}{0.57\textwidth}{(b) DNN$_1$ (left) and DNN$_2$ (right)}}
\caption{ResNet architectures for (a) single neural network and (b) parallel neural networks, where DNN$_1$ has output of mass fractions $X_k$ and DNN$_2$ nuclear energy generation $e_{\rm nuc}$. The solid lines represent standard ResNet shortcut connections while the dashed lines are additional connections. 
\label{fig:resnets}}
\end{figure}

All DNNs are implemented using a ResNet~\citep{he2016deep} architecture with differing number of hidden layers and shortcut connections. Although ResNets are most commonly used in classification problems instead of surrogate models, they have been shown to perform well for modeling problems where there are many inputs and few outputs. ResNets differ from fully connected neural networks in that they include additional connections between non-consecutive hidden layers, usually skipping two or more layers. In very deep networks, these ``shortcut" connections help to alleviate the vanishing gradient problem, which is when the gradients become increasingly small during backpropagation causing sub-optimal convergence of the first few layers of the network. After experimenting with varying the number of hidden layers, hidden nodes, and shortcut connections, we chose to use the architectures specified in Figure \ref{fig:resnets} for the DNNs that we are training and testing. Note that the single and parallel neural networks contain the same number of total hidden nodes. 

The activation function used in each hidden layer is the hyperbolic tangent activation function due to its derivatives being continuously differentiable. As a side note, the CELU activation function, which is also continuously differentiable, can be used to obtain faster convergence, but we found that it does not necessarily result in better accuracy in practice. Finally, a ReLU activation function is added to the output layer of DNN$_1$ to satisfy the physical constraint that mass fractions must be positive values. 

\subsection{Loss Function and Scaling}\label{sec:loss_function}

Since one of our chosen neural networks form a parallel ResNet architecture, each parallel DNN would use a separate loss function: one for the mass fractions and one for the nuclear energy generation. In the case of the single DNN, the total loss function would be the sum of the two parallel loss functions. Due to the fact that these neural networks are modeling systems of ODEs describing physical phenomena, there are many opportunities to incorporate physics constraints into the loss functions. In this particular problem, there are two such constraints we consider when constructing the loss function: 1) the sum of the mass fractions must be conserved in DNN$_1$, and 2) the nuclear energy generation must be of the same sign in DNN$_2$. Letting $\tilde{X}_k$ and $\tilde{e}_{\rm nuc}$ be the model-predicted values of the mass fractions and energy generation, and $X_k$ and $e_{\rm nuc}$ the solution states from the ODE solver, the loss functions for the two neural networks are
\begin{eqnarray}
\text{DNN}_1:& \qquad \mathcal{L}(\tilde{X}_k) = \sum_k \underbrace{
    \frac{1}{N}\sum_{i=1}^{N} \lVert\tilde{X}_k - X_k\rVert_2^2
    }_{\text{MSE}(\tilde{X}_k, X_k)} 
    + C_{X} \underbrace{
    \frac{1}{N}\sum_{i=1}^{N} \left\Vert\sum_k \tilde{X}_k - \sum_k X_k\right\Vert_2^2
    }_{\text{mass conservation constraint}} \\ 
\text{DNN}_2:& \qquad \mathcal{L}(\tilde{e}_{\rm nuc}) = \underbrace{
    \frac{1}{N}\sum_{i=1}^{N} \lVert\tilde{e}_{\rm nuc} - e_{\rm nuc}\rVert_2^2
    }_{\text{MSE}(\tilde{e}_{\rm nuc}, e_{\rm nuc})} 
    + C_{e} \underbrace{
    \frac{1}{N}\sum_{i=1}^{N} \lVert\text{sign}(\tilde{e}_{\rm nuc}) - \text{sign}(e_{\rm nuc})\rVert_1
    }_{\text{physical constraint}} \\[1em]
\text{DNN}_{\rm full}:& \qquad \mathcal{L}([\tilde{X}_k,\tilde{e}_{\rm nuc}]) = \mathcal{L}(\tilde{X}_k) + \mathcal{L}(\tilde{e}_{\rm nuc})
\end{eqnarray}
for data points $i = 1, \dots, N$, where $C_{X}$ and $C_{e}$ are cost factors for the physical constraints of mass fractions and energy generation, respectively. Here, MSE refers to the Mean Square Error between the ground truth solutions and the model-predicted values.

In order for neural networks to perform optimally, the inputs and outputs should be scaled such that they are close to order $\mathcal{O}(1)$. This is easily achievable for density, temperature, and nuclear energy generation, all of which are normalized with their respective maximum values. However, the mass fractions vary widely from $\mathcal{O}(10^{-30})$ to $\mathcal{O}(10^{-1})$ (see Figure \ref{fig:tn200_profile}) and hence cannot be normalized in a similar manner.
Instead, we employ two potential approaches to alleviate this scaling problem. The first is to convert all mass fractions to their inverse log equivalents, which will significantly reduce the range of scale. The second is to use a loss function with exponentially increasing weights that is symmetric around a mass fraction of 0.25, which is motivated by mass conservation. The results of using these two approaches will be discussed further in Section \ref{sec:Results}.

\begin{figure}[tb]
\centering
\includegraphics[width=0.45\linewidth]{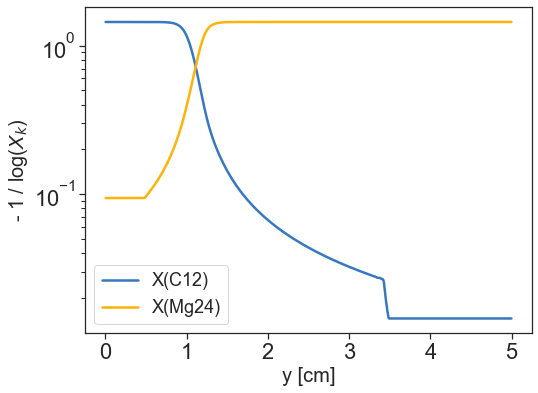}
\caption{Profile of mass fractions in negative inverse log form at $t=4.8\mu$s. 
\label{fig:log_Xk}}
\end{figure}

The first approach to addressing the mass fraction scaling problem is to transform both input and output mass fractions (in DNN$_1$) to their negative inverse log equivalent, $X_k \rightarrow -1/\log(X_k)$. This reduces the mass fractions to a much more narrow range of $\mathcal{O}(10^{-2})$ to $\mathcal{O}(1)$ as shown in Figure \ref{fig:log_Xk}. This transformation can be applied to the input mass fractions in DNN$_2$ as well.

\begin{figure}[tb]
\centering
\includegraphics[width=0.4\linewidth]{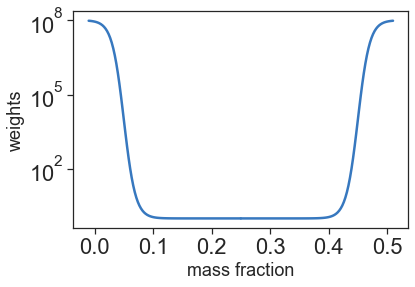}
\caption{Weights associated with the modified loss function of mass fractions. Note the symmetry around a mass fraction of 0.25.
\label{fig:loss_weights}}
\end{figure}

The second method is to note that in this particular 3-species network, the sum of $X_{\rm C12}$ and $X_{\rm Mg24}$ is conserved. Therefore, when either mass fraction approaches 0, the other must approach 0.5. In the previous approach where the inverse log form of mass fraction was used, we considered the scaling problem only near 0. However, due to the fact that the sum of the species is conserved, we realized that there is a scaling issue near 0.5 as well. In fact, we need a loss function that penalizes errors near 0 and 0.5 in a symmetric manner as well as take into account the scaling problem. This can be accomplished by using custom weights $w(X_k)$ in the loss function such that 
\begin{equation}
\mathcal{L}_{w}(\tilde{X}_k) = \frac{1}{N}\sum_{i=1}^{N} \sum_k w(X_k) \: \lVert\tilde{X}_k - X_k\rVert_2^2
\end{equation}
where the weight is a function of the mass fraction that is symmetric about $X_k=0.25$ and exponentially increases towards 0 and 0.5. More specifically, after some initial testing of the accuracy needed for the mass fractions during real-time simulation, the modified loss function must heavily penalize errors within the mass fraction ranges of $(0,0.1]$ and $[0.4,0.5)$. Therefore, the weight function that we selected is similar to a ``double" sigmoid function centered at $X_k = 0.05$ and $0.45$, and defined as
\begin{equation}
    \log_{10}\left\{w(X_k)\right\} = \begin{cases}
        -\dfrac{p}{\exp(-100X_k + 5) + 1} + p &\mbox{if } X_k < 0.25 \\[0.5em]
        \dfrac{p}{\exp(-100X_k + 5) + 1} &\mbox{if } X_k \ge 0.25
    \end{cases}
\end{equation}
where $p$ is the order of precision such that $10^p$ is the maximum value of the weight function. In other words, $p$ dictates the precision where the modified loss function starts to loosen penalties for any values less than order $\mathcal{O}(10^{-p})$.
Figure \ref{fig:loss_weights} shows an example of the weight function with $p=8$. During training, we set $p=10$, but based on some preliminary testing, any value of $p\ge 8$ gives stable results. 
Using this approach to solve the mass fraction scaling problem, the modified loss function for DNN$_1$ then becomes
\begin{equation}
\mathcal{L}(\tilde{X}_k) = \mathcal{L}_{w}(\tilde{X}_k) 
    + C_{X} \frac{1}{N}\sum_{i=1}^{N} \left\Vert\sum_k \tilde{X}_k - \sum_k X_k\right\Vert_2^2.
\end{equation}

\subsection{Training and Validation}\label{sec:training}
\begin{figure}[tb]
\centering
\plottwo{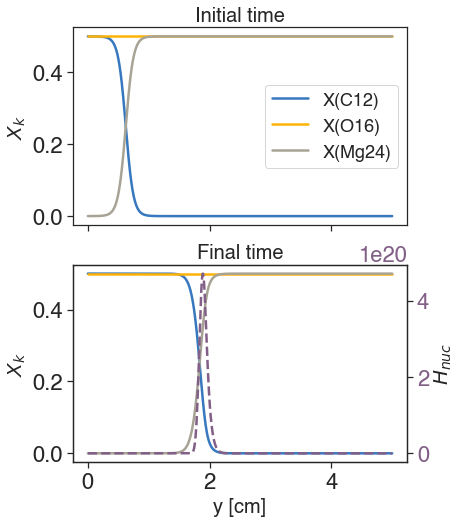}{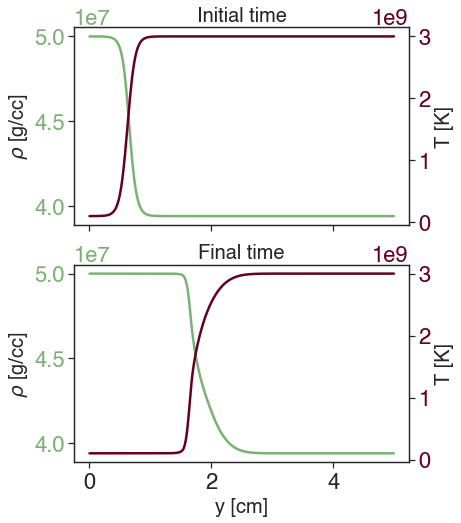}
\caption{Profile of species mass fractions, density, and temperature at initial time $t=0$ (top row) and final time $t = 12\mu$s (bottom row) for planar case.
\label{fig:igsimple_profiles}}
\end{figure}

We define initial conditions for our simulation as follows.
In each simulation, the computational domain is $L_x=0.625$~cm $\times$ $L_y=5$~cm with 128 $\times$ 1024 grid cells so the grid spacing is $\Delta x\approx 49~\mu$m. 
To initialize the problem, 
we define the fuel-state density ($\rho_{\rm fuel}=5\times 10^7$~g/cm$^3$), temperature ($T_{\rm fuel}=10^8$~K), and composition ($X_{\rm C12,fuel}=X_{\rm O16,fuel}=0.5$, $X_{\rm Mg24,fuel}=0$) and use the equation of state to compute the corresponding ambient pressure used throughout the domain, $p_0$.
We define the ash-state temperature ($T_{\rm ash}=3\times 10^9$~K) and composition ($X_{\rm C12,ash}=0$).
We define profiles for the temperature and species using a smooth hyperbolic tangent, 
\begin{equation}\label{eq:profile_T}
    T = T_{\rm fuel} + \left(\frac{T_{\rm ash}-T_{\rm fuel}}{2}\right) \left[1 + \tanh{(\alpha(y-\tilde{y}(x)) - 5)}\right],
\end{equation}
\begin{equation}\label{eq:profile_Xk}
    X_{\rm C12} = X_{\rm C12,fuel} + \left(\frac{X_{\rm C12,ash}-X_{\rm C12,fuel}}{2}\right) \left[1 + \tanh{(\alpha(y-\tilde{y}(x)) - 5)}\right],
\end{equation}
\begin{equation}
    X_{\rm Mg24} = 0.5 - X_{\rm C12}, \quad X_{\rm O16} = 0.5,
\end{equation}
where $\alpha=8$~cm$^{-1}$ is a parameter that controls the steepness of the transition.
Then we use the equation of state to compute $\rho$ and $h$ throughout the domain using the temperature, composition, and ambient pressure.
The term $\tilde{y}(x)$ represents a spatial perturbation to the initial flame front.
In our first tests, $\tilde{y}(x)=0$ so the flame front is on one plane (one-dimensional).
The x-boundary conditions are periodic, and the y-boundary conditions are inflow and outflow.
The inflow boundary conditions use a prescribed velocity of $\Ub=(0,10^5)$~cm/s with the fuel-state condition for the remaining variables.
The initial profiles of the species, temperature, and density for the one-dimensional case are shown in the top two panels in Figure \ref{fig:igsimple_profiles}.

Using the MAESTROeX algorithm with VODE, we model $12\mu$s of evolution over 500 time steps with $\Delta t=24$~ns, which corresponds to an advective CFL condition of $\sim$0.5.
The peak velocity in this simulation is very close to the inflow velocity, $10^5$~cm/s, which corresponds to a Mach number of $\mathcal{O}(10^{-4})$.
The final configuration of the species is shown in the bottom two panels in Figure \ref{fig:igsimple_profiles}.
Over this time, the burning front speed is much smaller than the inflow velocity, so the front travels $\sim$1~cm to the right, which is roughly 20\% of the length of the domain.  

\begin{figure}[tb]
\centering
\includegraphics[width=\linewidth]{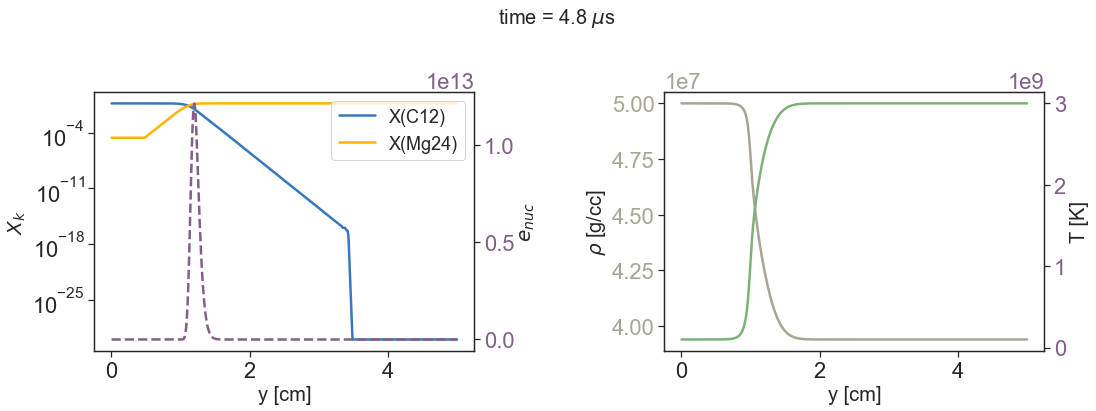}
\caption{Profiles of species mass fractions, density, and temperature at $t=4.8\mu$s, the time starting from when the training data is generated. 
\label{fig:tn200_profile}}
\end{figure}

The data we are using to train and validate the DNN models is based on reaction data from this simulation.
The inputs and outputs of the ODE solver VODE are saved to plotfiles that are then used to train the neural networks offline. These plotfiles are generated every 10 time steps starting from the time $t=4.8\mu$s to the final time of $t=12\mu$s. Figure \ref{fig:tn200_profile} shows the profile of the relevant variables at $t=4.8\mu$s.

\begin{figure}
\centering
\includegraphics[width=\linewidth]{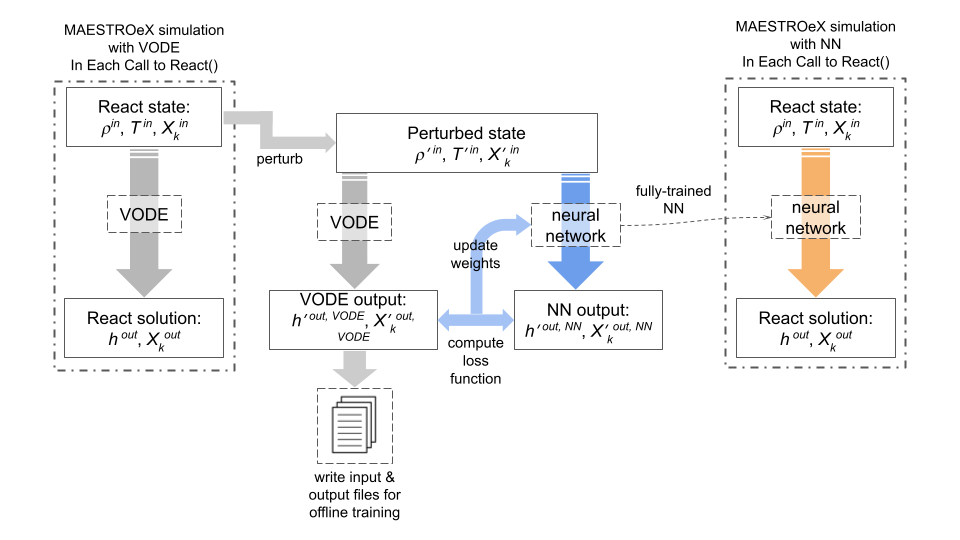}
\caption{Illustrated overview of our machine learning approach. The gray arrows indicate the steps involved in generating the training data, the blue arrows the training and validation process, and the orange arrows the testing phase.
\label{fig:flowchart}}
\end{figure}

We found in preliminary testing that simulations using neural networks that are trained using data directly from the MAESTROeX simulation tend to perform poorly for inputs that are slightly perturbed from the training data.
This indicates that the model is most likely overfitting since it only sees a small set of the possible solution space of the ODE solver. To improve robustness, stability, and overall performance of the DNNs, we present an approach to diversify the training data by slightly perturbing the fluid state just before using VODE to solve the system. The procedure to perturb the states is as follows.
\begin{enumerate}
    \item Apply a random perturbation within $\pm 0.005$ to $X_{\rm C12}$.
    \item Subtract the same perturbation from $X_{\rm Mg24}$ to ensure mass conservation. If any of the resulting mass fractions is negative, set the perturbation to zero (i.e.~do not perturb this cell).
    \item Randomly perturb both density and temperature within $\pm 0.02$\% and $\pm 0.06$\% respectively.
    \item Repeat steps 1 to 3 in 75\% of the cells in the computational domain.
\end{enumerate}
The overall process to generate the training data is shown by the gray arrows in Figure \ref{fig:flowchart}.  Note that we still run the actual simulation using unperturbed data; the perturbed data is passed in through separate, additional calls to VODE that do not feedback into the main simulation.  Also note that we complete training over a representative set of data over all time steps before using the trained model to start a DNN-based simulation. In future work we may explore training ``on-the-fly" to further improve the efficiency of the workflow.  In the one-dimensional case, the data set we chose is the solution along $x$-slices located at $x=[0.1, 0.2, 0.3]$~cm, and in the two-dimensional case, the training data is taken from half of the domain ($x\le 0.3125$~cm).

\begin{figure}[tb]
\centering
\includegraphics[width=0.45\textwidth]{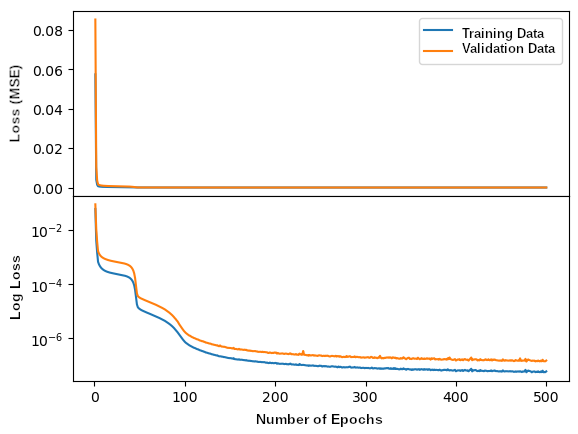}
\caption{Total loss of training and validation data for 500 epochs during validation phase.
\label{fig:cost_vs_epoch}}
\end{figure}

The validation data
is a subset of the training data that is not used to update the neural network but is instead used to measure the performance of the neural network during training time. The neural network makes predictions on this data to give an indicator of the error of the predictions to the ground truth, but does not actually back propagate gradients to optimize for this data.  Usually, a different loss function is used for validation. For simplicity, we chose to use the MSE as the validation loss function. We then split the total training data set into 10\% for validation and 90\% to be used to actually train the DNNs.  

In Figure \ref{fig:cost_vs_epoch} we show the training and validation losses during the first 500 epochs of training the full DNN model using negative inverse log scaling of the inputs. There are two important things to point out here. Firstly, the validation data loss is slightly higher, but follows a similar downward trend as the training data during training time. This is because the neural network has not optimized for this validation data, so it will never achieve the same loss as that of the training data; however, this gives us confidence that the model is learning. Secondly, the validation loss of this neural network has not diverged from the training loss. If the validation loss were to increase while the training loss continues to decrease, this would be evidence of overfitting. On the other hand, if the validation loss was simply not decreasing in the same manner as the training loss (i.e.~the validation loss plateaus at a higher loss cost), this would be evidence that our model is not complex enough to capture the intricacies of our data such that it generalizes well. This is a well known phenomena called the bias-variance trade-off. Here, we show our model is simple enough to generalize well in addition to being sophisticated enough to accurately capture the dynamics of the problem.

\subsection{Testing}\label{sec:testing}
The testing phase comes after training and validation of the neural networks.
In order to determine the best scaling approach and loss functions to use,
when running with trained DNN in place of VODE, we re-run the same flame diffusion simulation in Section \ref{sec:training} starting with the configuration at $t=4.8\mu$s to the final time of $t=12\mu$s.
After determining the best training approach, we train a new DNN with a perturbed flame with a V-shaped front, and test this model on three different configurations: the same V-shaped front, a one-dimensional flame, and a sinusoidally-varying flame front. The V-shaped front is defined using
\begin{equation}
\tilde{y}(x) = \begin{cases}
    -x &\mbox{if } x \le L_x/2 \\[0.5em]
    -L_x + x &\mbox{else} 
    \end{cases}
\end{equation}
where $\tilde{y}(x)$ is used in Equations (\ref{eq:profile_T}) and (\ref{eq:profile_Xk}) and again, $L_x=0.625$~cm is the width of the domain.
For our simulations involving a sinusoidally-varying flame front, we use a domain that is twice as wide (with twice as many grid cells in $x$ so that $\Delta x$ remains the same) and define the flame front using
\begin{equation}
\tilde{y}(x) = \frac{2L_{x}}{4}\sin\left(\frac{2\pi}{2L_{x}}x\right).
\end{equation}
The results of these simulations are discussed in detail in Section \ref{sec:Results}.

\subsection{Software Implementation}\label{sec:implementation}

The initial simulations used to generate the plotfiles that contain the training data is run using
MAESTROeX (\cite{fan2019maestroex,MAESTROeX_JOSS},
\url{https://github.com/AMReX-Astro/MAESTROeX} hash {\tt 717e4bc}),
which uses Starkiller Microphysics libraries (\cite{starkiller},
\url{https://github.com/AMReX-Astro/Microphysics}) tag {\tt 22.07}),
and the AMReX framework (\cite{zhang2021amrex,AMReX,AMReX_JOSS},
\url{https://github.com/AMReX-Codes/amrex}) tag {\tt 22.07}).

The training and validation of the neural networks are implemented in Python using the PyTorch library~\citep{pytorch} and yt~\citep{yt} is used to extract the training data from the plotfiles. 
The main reason we chose to use PyTorch over other machine learning libraries is due to the simplicity of its C++ API LibTorch for integration into our existing C++ codes. By using PyTorch's tracing JIT and LibTorch library, we are able to export the trained neural networks to a model file that can then be imported and used in C++. The integration of LibTorch into the AMReX framework allows the user to take advantage of AMReX's parallelism capability on both host (MPI+OpenMP) and MPI+GPU. We have implemented the use of the LibTorch library in AMReX and, by extension, MAESTROeX in order to test the trained neural networks. The problem set-up is found in \texttt{/Exec/science/flame\_ml/} directory of MAESTROeX.

\section{Results}\label{sec:Results}
As outlined in Section \ref{sec:testing}, we will first analyze the performance of the full DNN and parallel DNNs by training and testing with the same one-dimensional planar problem. Then after determining the best model to use, we will train a new model using data from a flame with a V-shaped front and test it on three different configurations: the same V-shaped front, a one-dimensional flame, and a sinusoidally-varying flame front. For each of the test cases, the final solution at $t=12\mu$s is compared between using VODE and the trained DNN. In addition, plots of the model prediction versus the ground truth solution of the output variables are included to show how well the model performs on the testing data at the end of the training phase. The closer the points are to the $y=x$ line, the closer the prediction is to the ground truth.

\subsection{Full DNN}\label{subsec:result_full}
\begin{figure}[tb]
\gridline{\fig{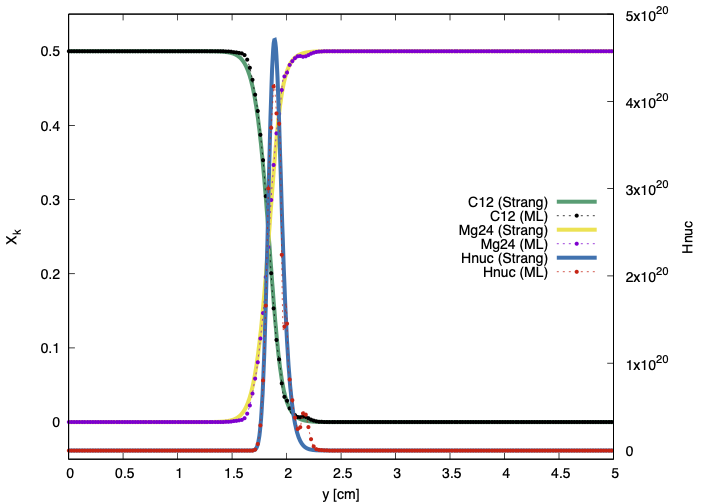}{0.52\textwidth}{(a)}
          \fig{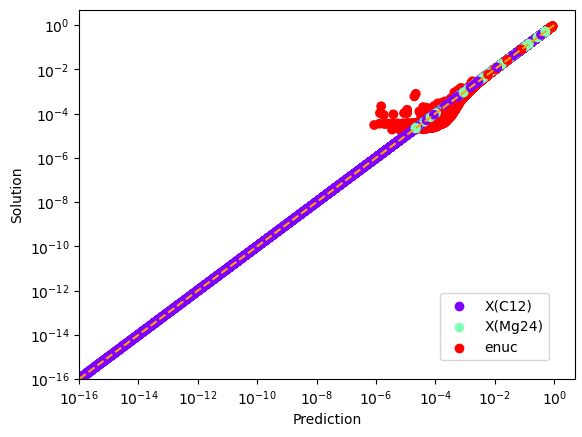}{0.46\textwidth}{(b)}}
\caption{(a) Profile of species mass fractions and nuclear energy generation rate at final time $t = 12\mu$s, and (b) validation results using the negative inverse log form of $X_k$ as input for full DNN. Note that each point in the profile plot is every 5 data points on the grid in the $y$-direction.
\label{fig:result_full_Xlog}}
\end{figure}

\begin{figure}[tb]
\gridline{\fig{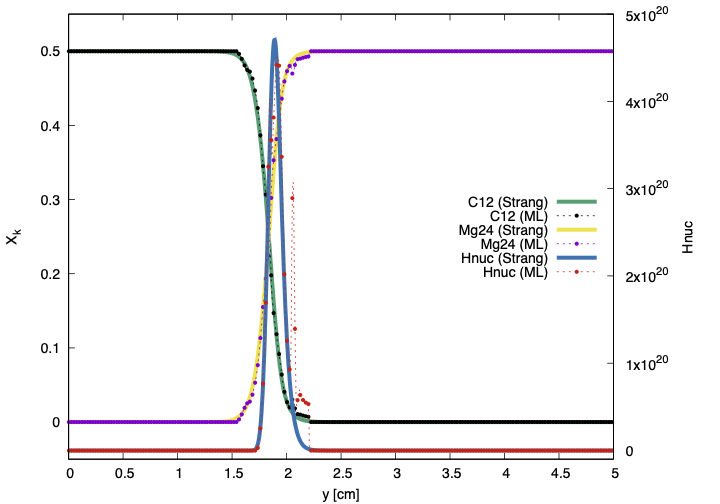}{0.52\textwidth}{(a)}
          \fig{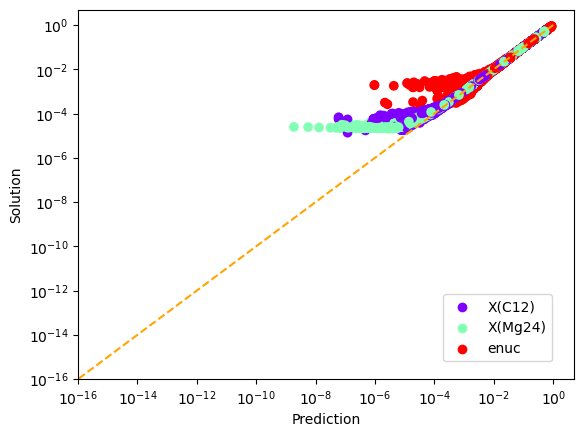}{0.46\textwidth}{(b)}}
\caption{(a) Profile of species mass fractions and nuclear energy generation rate at final time $t = 12\mu$s, and (b) validation results using symmetric mass fraction loss function for full DNN.
\label{fig:result_full_wexp}}
\end{figure}

The full DNN was trained using data from the one-dimensional planar flame problem for 3000 epochs, and the results are shown in Figures \ref{fig:result_full_Xlog} and \ref{fig:result_full_wexp}. Of the two scaling strategies discussed in Section \ref{sec:loss_function}, the negative inverse log transformation of $X_k$ inputs seems to be the better choice, although both neural networks show similar shortcomings in performance. The first issue is the emergence of oscillations at the front of the flame, which suggests that the solution has become unstable. The oscillations are clearly more pronounced when using the symmetric mass fraction loss function. The second issue is that both full DNN models have underestimated the peak nuclear energy generation rate by a noticeable margin. In this case, the log transformation of $X_k$ inputs underestimated the peak by $\sim$11\% while the symmetric loss function underestimated by $\sim$4\%. 

\subsection{Parallel DNN}\label{subsec:result_parallel}
\begin{figure}[tb]
\centering
\sbox{0}{\includegraphics[width=0.65\textwidth]{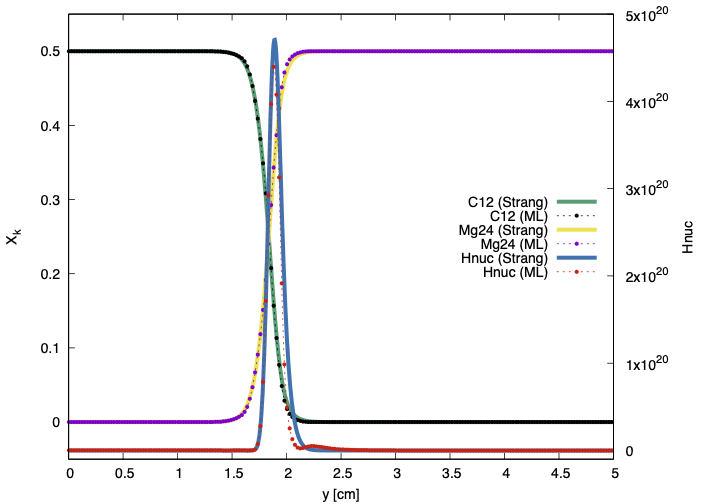}}
\begin{tabular}{@{}c@{}c@{}}
\usebox{0} &
  \parbox[b][\ht0][s]{0.3\textwidth}{
    \includegraphics[width=0.3\textwidth]{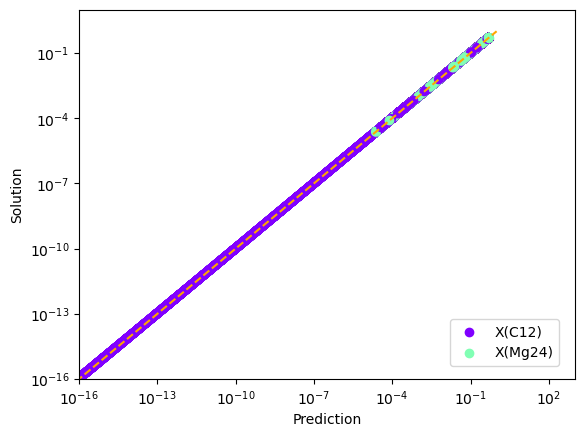}
    \vfill
    \includegraphics[width=0.3\textwidth]{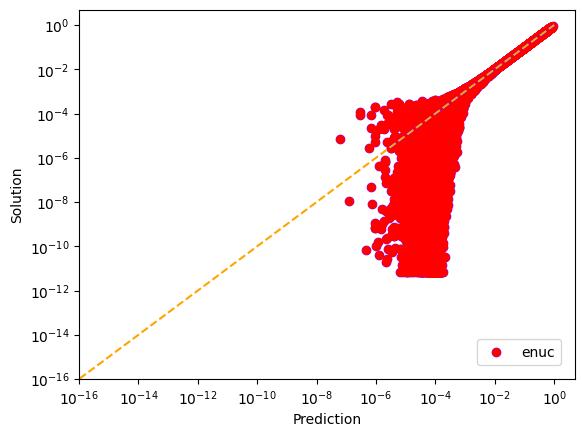}
  }
\end{tabular}
\caption{(left) Final profiles of mass fractions and nuclear energy generation rate at $t=12\mu$s,
and (right) parallel validation results using the negative inverse log form of $X_k$ as input for both DNNs. Note that each point in the profile plot is every 5 data points on the grid in the $y$-direction.
\label{fig:result_parallel_logX}}
\end{figure}

\begin{figure}[tb]
\centering
\sbox{0}{\includegraphics[width=0.65\textwidth]{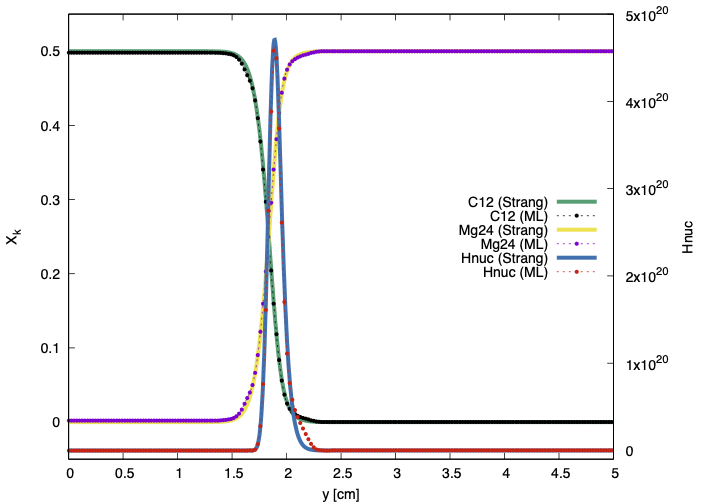}}
\begin{tabular}{@{}c@{}c@{}}
\usebox{0} &
  \parbox[b][\ht0][s]{0.3\textwidth}{
    \includegraphics[width=0.3\textwidth]{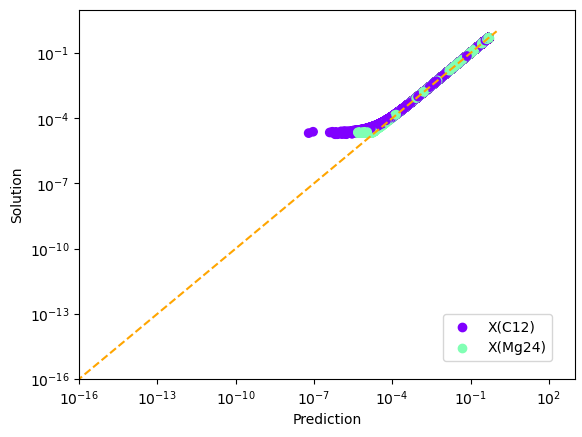}
    \vfill
    \includegraphics[width=0.3\textwidth]{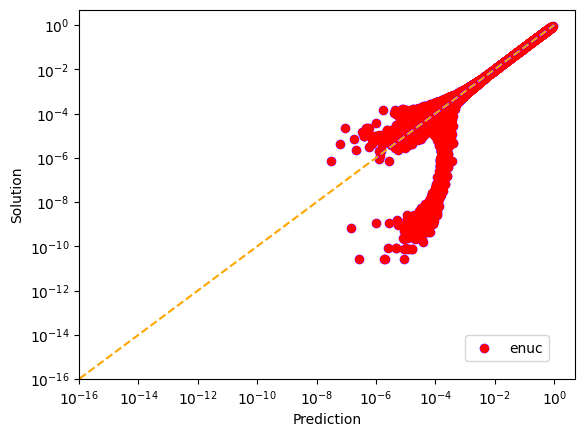}
  }
\end{tabular}
\caption{Final profiles of mass fractions and nuclear energy generation rate at $t=12\mu$s (left),
and parallel validation results (right) using symmetric mass fraction loss function for DNN$_1$.
\label{fig:result_parallel_wexp}}
\end{figure}

Similar to the full DNN, each parallel DNN was trained using data from the one-dimensional planar flame problem for 3000 epochs, and the results are shown in Figures \ref{fig:result_parallel_logX} and \ref{fig:result_parallel_wexp}. From these results, we can easily conclude that the parallel DNN performed much better than the full DNN. There is no longer any discernible oscillations at the front of the flame and the profiles of all three neural network outputs (dotted lines) are very similar to the original Strang solution using VODE (solid lines). However, the scaling strategy using the negative inverse log transformation of $X_k$ inputs underestimated the peak nuclear energy generation rate by $\sim$6\% while the symmetric mass fraction loss function only underestimated the peak by $<$1\%.

The improvement in performance of the parallel DNNs over the full DNN is likely due to the fact that the loss function of the mass fractions can be derived independently from that of the nuclear energy generation. By defining completely separate neural networks for the mass fractions and $e_{\rm nuc}$, the resulting parallel neural networks are able to model each output state with higher accuracy. Interestingly, the time it takes to train one epoch of data on the parallel neural network is only 4\% faster than the full neural network, which is much less of a speedup than expected. This is most likely caused by the uneven distribution of nodes and hidden layers in DNN$_1$ and DNN$_2$ with the former containing two-thirds of the total number of hidden nodes and much wider hidden layers than the latter.
See Appendix \ref{app:enuc_analysis} for a more detailed discussion on the use of the parallel DNN for separating the mass fraction and $e_{\rm nuc}$ neural networks.

\subsection{Two-dimensional Flame Problems}\label{subsec:result_2d}
\begin{figure}[tb]
\gridline{\fig{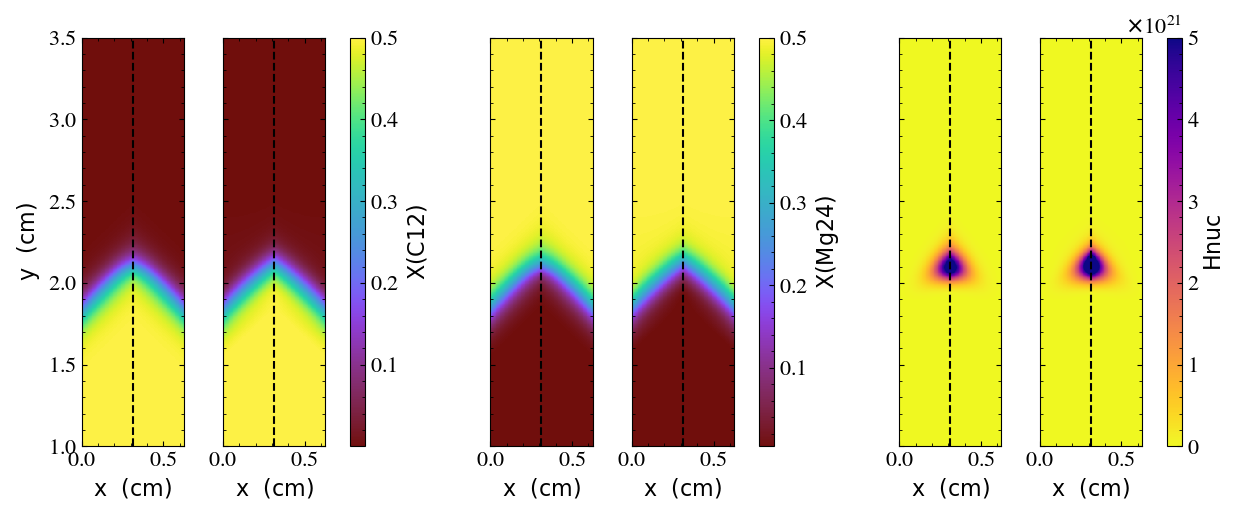}{0.8\textwidth}
            {(a) Contour plots of species mass fractions and nuclear energy generation rate. Each pair of contour plots shows the Strang solution using VODE on the left and trained DNN model on the right along the vertical dashed line.}}
\gridline{\fig{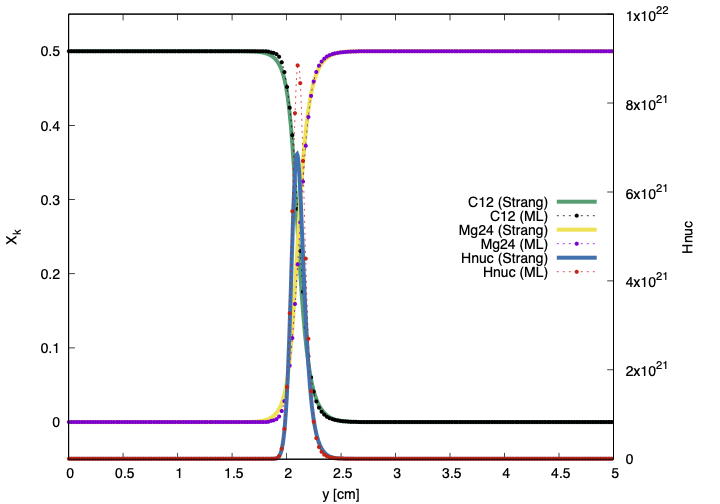}{0.7\textwidth}
            {(b) Profiles along the vertical dashed line.}}
\caption{Solutions to the two-dimensional V-shape problem at final time $t = 12\mu$s. Note that each point in the profile plot is every 5 data points on the grid in the $y$-direction.
\label{fig:flame_vshape}}
\end{figure}

Based on the results in Sections \ref{subsec:result_full} and \ref{subsec:result_parallel}, we chose to train the parallel neural network using the symmetric mass fraction loss function on the two-dimensional flame with a V-shaped front problem. The DNNs were trained using data from half of the domain ($x\le 0.3125$~cm) for 500 epochs, and the results at the final time are shown in Figure \ref{fig:flame_vshape}. The mass fraction solutions and location of the flame front using the parallel DNN are comparable to those using VODE. However, the peak $H_{\rm nuc}$ prediction has an overshoot of approximately 28\%. One possible explanation for this subpar performance is that the solution space of the ODE near the peak of the V-shaped flame front make up only a small portion of the training data set. This is especially true near the end of the simulation as the peak $H_{\rm nuc}$ increases over time and would only exist in the last couple of plotfiles used to generate the training data set. Hence, because the neural network does not see as many instances of the solution near the peak $H_{\rm nuc}$, it is possible that the model does not have enough information to make good predictions there. 

\begin{figure}[tb]
\gridline{\fig{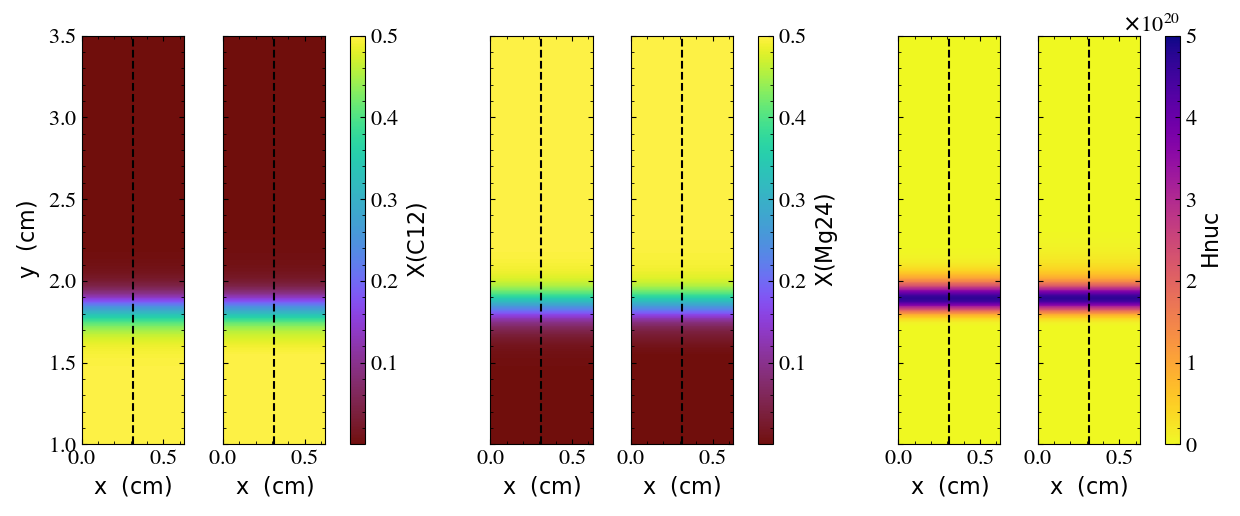}{0.8\textwidth}
            {(a) Contour plots of species mass fractions and nuclear energy generation rate. Each pair of contour plots shows the Strang solution using VODE on the left and trained DNN model on the right along the vertical dashed line.}}
\gridline{\fig{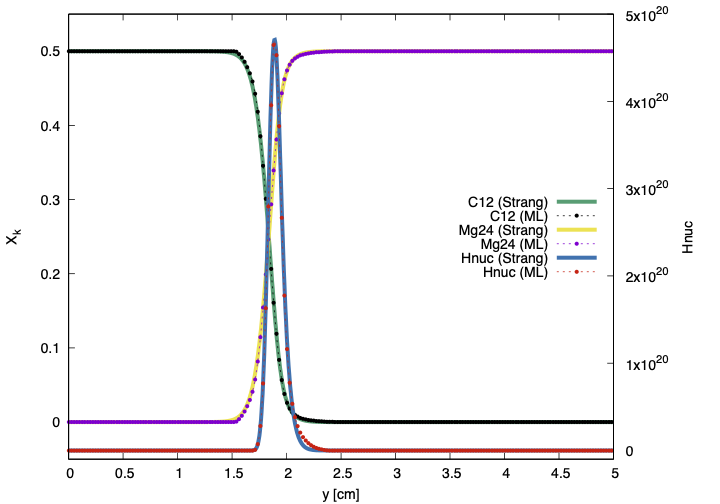}{0.7\textwidth}
            {(b) Profiles along the vertical dashed line.}}
\caption{Solutions to the planar problem at final time $t = 12\mu$s using an ML model trained using data from the V-shape problem.
\label{fig:flame_planar}}
\end{figure}

By contrast, when we tested this trained DNN on the one-dimensional planar problem, the solution at peak $H_{\rm nuc}$ is almost indistinguishable from the VODE solution as shown in Figure \ref{fig:flame_planar}. Again, this could be attributed to the fact that there are many more instances of solutions closer to those encountered in the planar problem in the training data set and as a result the neural network is able to produce much more accurate predictions. Given these results, a potential solution to improve the performance of the DNN for the V-shaped flame problem is to include more instances of the flame near the peak $H_{\rm nuc}$ either by expanding the training data set to include data from two-third of the domain ($x\le 0.4167$~cm) or to output multiple plotfiles with varying solution perturbations (as described in Section \ref{sec:training}) near the end of the simulation.

\begin{figure}[tb]
\gridline{\fig{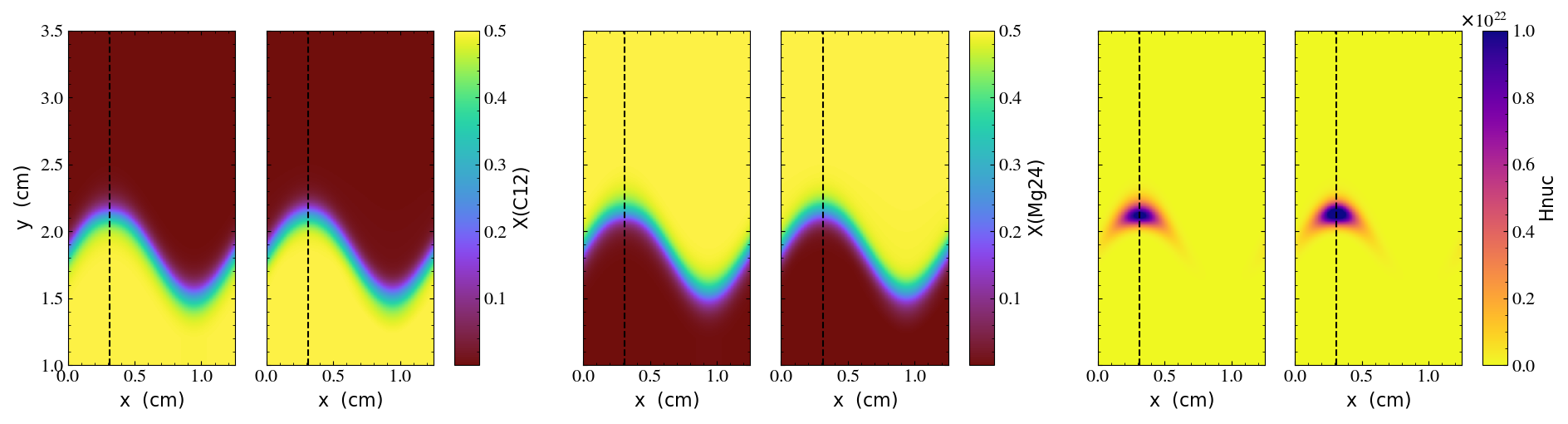}{\textwidth}
            {(a) Contour plots of species mass fractions and nuclear energy generation rate. Each pair of contour plots shows the Strang solution using VODE on the left and trained DNN model on the right along the vertical dashed line.}}
\gridline{\fig{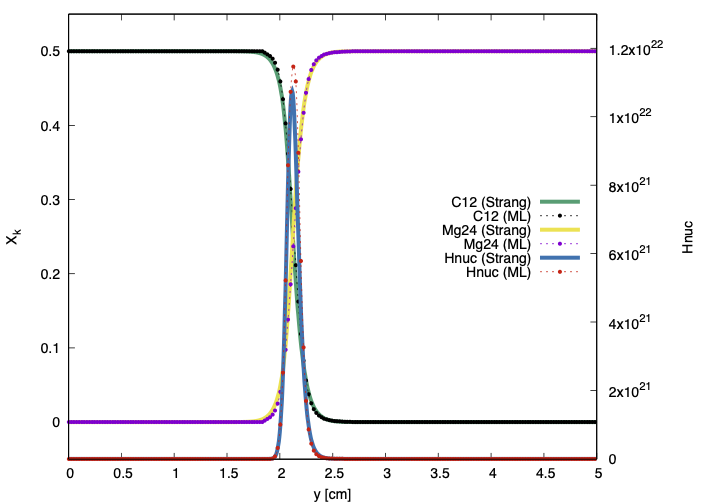}{0.7\textwidth}
            {(b) Profiles along the vertical line.}}
\caption{Solutions to the sine wave problem at final time $t = 12\mu$s using an ML model trained using data from the V-shape problem.
\label{fig:flame_sine}}
\end{figure}

As the final test, we test the trained parallel DNN on another two-dimensional problem that differs from the training set data, which is a flame with a sinusoidally-varying flame front.
In this problem, the computational domain is $1.25\text{cm} \times 5\text{cm}$ with $256 \times 1024$ grid cells so that the grid spacing stays the same as in the previous problems. Figure \ref{fig:flame_sine} shows that the neural network solutions of both mass fractions and nuclear energy generation rate are close to the VODE solutions, with a slight overshoot in peak $H_{\rm nuc}$ of approximately 7\%. 

\subsection{Timing Comparisons}\label{subsec:result_timing}
\begin{figure}[tb]
\centering
\includegraphics[width=0.99\textwidth]{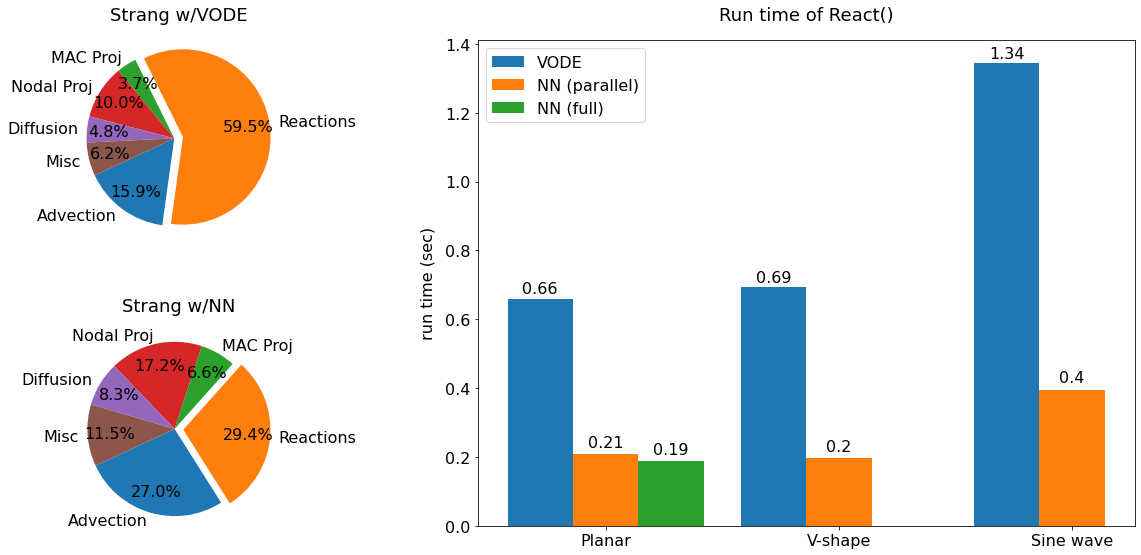}
\caption{(left) Relative timing plots of subroutines in MAESTROeX. (right) Comparison of run times of the computational reaction kernel using VODE and DNN models.
\label{fig:result_timings}}
\end{figure}

To compare the runtime performance of MAESTROeX when using a neural network instead of the ODE solver VODE for the computational reaction kernel, we run the three previously defined flame problems (planar, V-shaped, and sinusoidally-varying flame fronts) on four cores (pure MPI) on an AMD EPYC 7702P processor and record the time it takes to complete the reactions subroutine per time step. The left plots in Figure \ref{fig:result_timings} show that in our simulations using VODE, the runtime is dominated by reactions, using 59.5\% of total runtime, whereas using DNN reduced the reactions runtime to 29.4\%. In terms of the time it takes to run the reaction kernels per time step, using the neural network results in speedups of 3.14 to 3.45 depending on the test problem (see right plot in Figure \ref{fig:result_timings}). 

\section{Conclusions and future work}\label{sec:Conclusions}
We have demonstrated that we can train deep neural networks with a ResNet architecture using data obtained from the ODE solver VODE given that the data have been perturbed slightly to represent a larger portion of the possible solution space of the solver. In terms of accuracy and stability during simulation using MAESTROeX, the parallel DNN showed much better performance than the full DNN. In addition, the trained neural network can be used successfully in a test case that it has never seen before (i.e.~a flame with a sinusoidally-varying front) given that the training data was obtained from a similar test (i.e.~a flame with a V-shaped front). We also presented two strategies to mitigate the mass fraction scaling problem. Of the two strategies, the one that uses the symmetric mass fraction loss function performed slightly better than transforming the mass fraction inputs to its negative inverse log form. 

\begin{figure}[tb]
\centering
\includegraphics[width=0.65\textwidth]{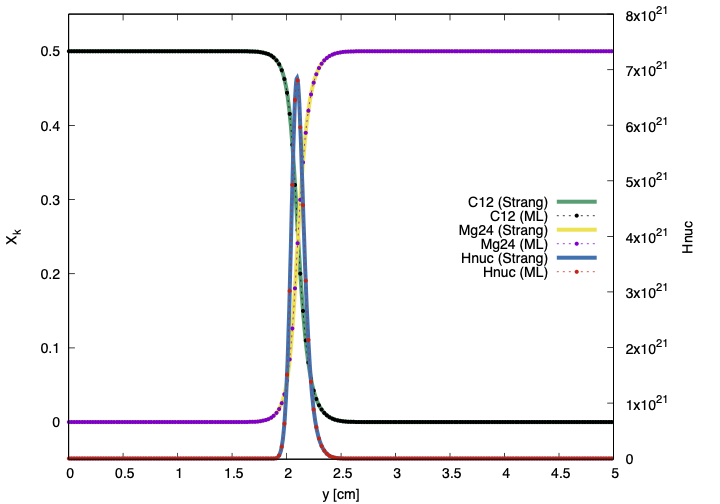}
\caption{Profiles of mass fractions and nuclear energy generation rate along the $x$-center line for the V-shaped flame problem at final time $t=12\mu$s using a parallel DNN in the predictor step of second-order Strang splitting algorithm.
\label{fig:result_predictor}}
\end{figure}

One area of potential future work that looks promising is using a neural network model to compute the reaction solution in the predictor steps of a high-order numerical algorithm such as Spectral Deferred Corrections (SDC) method. Even in the second-order Strang splitting algorithm that MAESTROeX currently uses, a slight speedup of 1.32 over all reactions steps can be seen if we replace the reaction ODE solve in the predictor step with our trained DNN. We would expect to see larger speedups in high-order methods that consists of multiple predictor steps. In addition, when we applied this approach in the two-dimensional flame with a V-shape front problem using the same parallel DNN that we previously trained in Section \ref{subsec:result_2d}, the final solution is exceedingly close ($<0.2$\% error) to the one from the original Strang splitting algorithm (see Figure \ref{fig:result_predictor}). 

Another topic for future work could be to develop and train neural networks with an additional input of the time step, which would lead to potential implementation of PINNs. However, in the case of stiff ODEs like the ones describing the nuclear reaction networks used in MAESTROeX, we would most likely need to consider a relaxation constraint instead of directly computing the gradients in the ODE~\citep{ji2020stiff}. Finally, we could look to model more complex nuclear reaction networks with 13 or 19 isotopes in the future. Modeling more complex networks using neural networks should result in larger speedups on both a CPU and GPU.

\begin{acknowledgments}
The work at LBNL was supported by the U.S. Department of Energy’s Scientific Discovery Through Advanced Computing (SciDAC) program under contract No.~DE-AC02-05CH11231.
The work at Stony Brook was supported by DOE/Office of Nuclear Physics grant DE-FG02-87ER40317.
This research used resources of the National Energy Research Scientific Computing Center (NERSC), a U.S. Department of Energy Office of Science User Facility operated under Contract No.~DE-AC02-05CH11231.
The authors thank Alice Harpole for valuable contributions to the Starkiller Microphysics library.
\end{acknowledgments}

%

\vspace{5mm}
\facilities{NERSC}


\software{Refer to Section \ref{sec:implementation}.}

\appendix
\section{Nuclear energy generation sensitivity analysis}\label{app:enuc_analysis}

Theoretically, the enthalpy could be updated using only the mass fractions. This is because after the evolving the mass fractions from ${\bf X}^{\rm in} \rightarrow {\bf X}^{\rm out}$, the resulting reaction rates are computed as 
\begin{equation}\label{eq:omega_out}
(\rho\dot{\omega}_k)^{\rm out} = \frac{\rho^{\rm out}(X_k^{\rm out} - X_k^{\rm in})}{\Delta t}
\end{equation}
and the nuclear energy generation rate becomes 
\begin{equation}\label{eq:Hnuc_sum}
(\rho H_{\rm nuc}) = -\sum_k (\rho\dot{\omega}_k)^{\rm out} q_k.
\end{equation}
Substituting Equations (\ref{eq:omega_out}) and (\ref{eq:Hnuc_sum}) into the simplest discrete form of Equation (\ref{eq:pde_h}) gives
\begin{eqnarray}
(\rho h)^{\rm out} &=& (\rho h)^{\rm in} - \Delta t\sum_k (\rho\dot{\omega}_k)^{\rm out} q_k \nonumber \\
 &=& (\rho h)^{\rm in} + \rho^{\rm out} e_{\rm nuc} \label{eq:discrete_h}
\end{eqnarray}
where $e_{\rm nuc}$ is a function of $({\bf X}^{\rm in}, {\bf X}^{\rm out}, q_k)$ and satisfies 
$$
H_{\rm nuc} = \frac{\partial e_{\rm nuc}}{\partial t}.
$$
Comparing Equations (\ref{eq:dnn_h}) and (\ref{eq:discrete_h}), it can be easily seen that $\text{DNN}_2$ is used to model $e_{\rm nuc}$. The question then is whether this additional DNN$_2$ is necessary given that $e_{\rm nuc}$ can be computed directly using the updated mass fractions from DNN$_1$.

\begin{figure}
\gridline{\fig{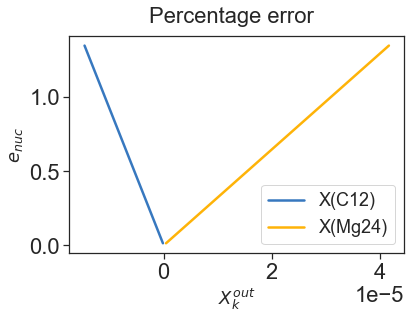}{0.42\textwidth}{\begin{eqnarray*}
            (X_{\rm C12}^{\rm in}, X_{\rm O16}^{\rm in}, X_{\rm Mg24}^{\rm in}) &=& (0.1483, 0.5, 0.3517) \\
            \rho^{\rm in} &=& 4.2996\times 10^7\text{ g cm}^{-3} \\
            T^{\rm in} &=& 2.2745\times 10^9\text{ K}
            \end{eqnarray*}}
          \fig{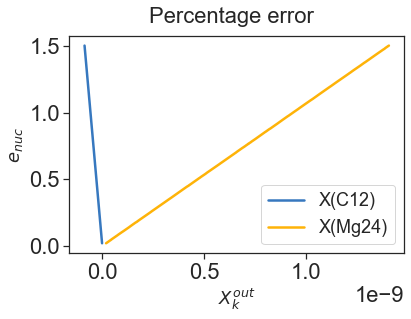}{0.42\textwidth}{\begin{eqnarray*}
            (X_{\rm C12}^{\rm in}, X_{\rm O16}^{\rm in}, X_{\rm Mg24}^{\rm in}) &=& (0.4381, 0.5, 0.06195) \\
            \rho^{\rm in} &=& 4.7228\times 10^7\text{ g cm}^{-3} \\
            T^{\rm in} &=& 3.2572\times 10^9\text{ K}
            \end{eqnarray*}}}
\caption{Percentage error of nuclear energy generation $e_{\rm nuc}$ with respect to percentage error of updated mass fractions $X_k^{\rm out}$ at different initial conditions.
\label{fig:enuc_sensitivity}}
\end{figure}

To answer this question, we want to analyze the sensitivity of $e_{\rm nuc}$ with respect to changes in the updated mass fractions (i.e.~the accuracy of DNN$_1$). Figure \ref{fig:enuc_sensitivity} plots the error in $e_{\rm nuc}$ relative to the errors in $X_k^{\rm out}$ at two different initial conditions: (left) at peak $e_{\rm nuc}$ of the flame and (right) approximately 0.2 cm behind the peak. Note that due to mass conservation, any change in $X_{\rm C12}$ will result in the same exact (but opposite) change in $X_{\rm Mg24}$. Figure \ref{fig:enuc_sensitivity}(a) shows that at peak energy generation of $6.0954\times 10^{12}$ erg/g, to achieve $\sim$99\% accuracy for $e_{\rm nuc}$ (i.e.~$\sim$1\% error) requires an error of $\sim 3\times10^{-5}$\% for $X_{\rm Mg24}$ and $\sim 1.5\times10^{-5}$\% for $X_{\rm C12}$. Even worse, with the initial conditions shown in Figure \ref{fig:enuc_sensitivity}(b), achieving a $\sim$99\% accuracy for $e_{\rm nuc}$ requires errors of $\sim 1\times10^{-9}$\% and $\sim 1\times10^{-10}$\% for $X_{\rm Mg24}$ and $X_{\rm C12}$, respectively. In practice, obtaining such high precision for neural networks is difficult and potentially unreasonably expensive due to memory limitations and longer training times. Therefore, we conclude that a second DNN is necessary to efficiently model the nuclear energy generation.

\bibliography{ml_reactions}
\bibliographystyle{aasjournal}

\end{document}